\documentclass[amssymb,aps,prl,twocolumn,groupedaddress,showpacs]{revtex4}
\usepackage{epsfig}
\usepackage{graphicx}
\usepackage{amssymb}
\usepackage{bm}
\usepackage{subfigure}

\begin{document}
\title{
Solid state effects on exciton states and optical properties of PPV}

\author{Alice Ruini,$^{1,2}$ Marilia J. Caldas,$^{3,1}$
Giovanni Bussi,$^{1,2}$ and Elisa Molinari$^{1,2}$ }

\affiliation{$^1$Istituto Nazionale per la Fisica della Materia (INFM)\\
$^2$Dipartimento di Fisica, Universit\`a di Modena e Reggio Emilia,
Via Campi 213a, I-41100 Modena, Italy \\
$^3$Instituto de Fisica, Universidade de Sao Paulo,
CP 66318, 05315-970 Sao Paulo, Brazil }

\date{\today}

\begin{abstract}
We perform {\it ab initio} calculations of optical properties 
for a typical semiconductor conjugated polymer, 
poly-{\em para}-phenylenevinylene, 
in both isolated chain and crystalline packing. 
In order to obtain results for excitonic energies and real-space wavefunctions
we explicitly include electron-hole 
interaction within the density-matrix formalism.
We find that the details of crystalline arrangement crucially affect
the optical properties, leading to a richer exciton structure and 
opening non-radiative decay channels.
This has implications for the 
optical activity and optoelectronic applications of polymer films.  
\end{abstract}


\maketitle
Ordered films of organic conjugated polymers are of  
strategic relevance for novel optoelectronic devices \cite{frie+99nat}. In
addition, they offer an ideal scenario for the study of electronic and 
excitonic confinement, being composed of quasi-one-dimensional (1D)
systems, arranged however in a three-dimensional (3D) crystalline environment. 
Long linear chains can indeed be thought of as 1D systems
with the highest degree of 1D confinement for a polyatomic
system ($\sim$ 2-5 \AA). 
Since the main optoelectronic characteristics derive from the mobile
$\pi$-electrons, delocalized along the chain backbone, and non bonding
to neighbouring chains,  the vast majority of studies  adopt the
single-chain model: in fact, complete {\it ab initio} theoretical studies of
the optical properties have been performed for 
isolated chains, highlighting the strong confinement expected for such
systems \cite{rohl-loui99prl,ruin+01sm}. Recently, very simplified
models of polymers in a 
``crystalline medium'' have been approached through {\it ab initio}
theory \cite{hors+99prl};  
however  the effect of crystal structure or side chains is completely
neglected. This is also usual 
practice in the intepretation of experimental data: e.g. data on 
poly-{\em para}-phenylenevinylene (PPV) and its alkylated and
metoxilated derivatives (MH-PPV, MEH-PPV, etc\dots) are usually bundled
together as representative of PPV, based on the supposition that
details of the 3D structure 
are not relevant.  The picture emerging from this analysis is far from
clear, however: 
quoted exciton binding energies differ by an order of 
magnitude \cite{leng+94prl,chan+97prb,alva98-01all,camp+96prl,
bart-bass97prl,mose+01sm_a}, there
is an on-going controversy about the existence of charge-transfer excitons 
or excimers \cite{conw96sm,wu-conw97prbrc,yan+95prl,camp+96prl,frol+00prl}, 
and about conditions for efficient light emission. 
The question then arises if, on the contrary, solid state effects
cannot be negleted and crystallization entails
also structure-specific 3D interchain coupling effects: recent
data on oligothiophene crystals seem to suggest that 
this might be the case \cite{mucc+00prb,buss+02tobe}.
 
Here we address these issues through a full {\it ab initio}
calculation of optical spectra and real-space exciton wavefunctions for
PPV, in both the isolated chain and the crystalline phase. In the
single PPV chain, we find that 
the lowest singlet exciton extends over a few monomers along the chain and
is optically active, with a binding energy of about 0.6$\div$0.7 eV.
In the crystal we find evidence of significant interchain coupling,
with decrease of the first optically active exciton binding energy
to about 0.2$\div$0.3 eV \cite{alva98-01all,camp+96prl,bart-bass97prl}; 
as a consequence of both interchain interaction and detailed
crystalline structure, a dark split-off singlet exciton appears 
at lower energy, which introduces a new path for non-radiative
emission and electron-hole dissociation, and is expected to decrease
drastically the photoluminescence efficiency \cite{yan+95prl};
furthermore, we find a much richer exciton structure, with charge-transfer
excitons close enough to the first direct exciton \cite{camp+96prl} to account 
for early electron-hole dissociation and apparently-vanishing 
binding energies \cite{mose+01sm_a};
fitting all the pieces together, our results can 
explain the controversies on the optical behavior of these crystals. 

The exciton binding energy, $E_b$, is a central quantity
in the photophysics of polymer semiconductors because it is related to the
spatial extension of the excitation and to the probability of radiative
emission/absorption and electric-field-induced dissociation of the exciton.
It is defined as the energy required to separate a bound electron-hole pair,
the exciton, into a free electron and a free 
hole \cite{conw96sm,alva98-01all,camp+96prl,bart-bass97prl,hill+00cpl}. 
For PPV, the
experimental values that are currently reported range from 
0.2$\div$0.4 eV \cite{alva98-01all,camp+96prl,bart-bass97prl} to around
1 eV \cite{leng+94prl,chan+97prb}, 
but energies of less than 0.1 eV are also proposed \cite{mose+01sm_a}.
These values correspond to spatial extensions varying from a single monomer
(Frenkel-like excitons) up to many unit cells (Wannier-like excitons).

From the theoretical point of view, calculating the properties of excitons in
extended polymeric systems is much more difficult than in conventional
inorganic-semiconductor quantum wires \cite{ross-moli96all}, 
because it requires to
combine an accurate treatment of the Coulomb interaction with a realistic
microscopic description of the single-particle electronic structure
beyond effective mass approximations.  Standard quantum-chemistry
schemes \cite{belj+99jcp} for treating 
excitations in finite molecular systems are intrinsically
not suitable for crystals. 
To this end, we have implemented a first-principles density-matrix approach
where the correlated electron-hole spectrum is obtained directly from the
interband polarizations $\langle d_{\nu}^{\dagger}c_{\mu}^{\dagger} \rangle $,
resulting from the creation of an electron in conduction
state $\mu$, with energy $\epsilon^{e}_{\mu}$ 
and a hole in valence state $\nu$ with energy $\epsilon^{h}_{\nu}$ 
($c^{\dagger}_{\mu}$ and $d^{\dagger}_{\nu}$ are the Fermionic creation 
operators; the labels $\mu = (c,{\bf k})$ and $\nu = (v,{\bf k})$ stand 
for both the band index and the wavevector). 
The polarization eigenstates are obtained by direct
diagonalization of the two-body Schr\"odinger equation:
$$
(\bar{\epsilon}^{e}_{\mu} - \bar{\epsilon}^{h}_{\nu})  A_{\mu\nu} +
e^2 \!\!\sum_{{\mu'}{\nu'}} 
(2\delta_S V_{\mu\nu',\nu\mu'} - W_{\mu\mu',\nu\nu'}) 
A_{\mu'\nu'}
\! = E_x A_{\mu\nu} 
$$
Here the first term on the right-hand side accounts for the uncorrelated
electron and hole quasi-particle energies, the second for electron-hole
screened direct ($W$) and unscreened exchange ($V$) Coulomb interaction
($\delta_S = 1$ for singlets, 0 otherwise).  
The interaction kernels are written in terms
of the single-particle wavefunctions $\psi$:
$$
\!\!\!\!W_{\mu\mu'\!,\!\nu\nu'}\!=\!\!\!
\int\!\!  \psi^{e *}_{\mu}({\bf r}_1)
\psi^{e}_{\mu'}({\bf r}_1) \
\!\! w({\bf r}_1 , {\bf r}_2 )
\psi^{h *}_{\nu}({\bf r}_2) \psi^{h}_{\nu'}({\bf r}_2)
 d{\bf r}_1 d{\bf r}_2
$$
\vspace{-0.6truecm} 
$$
 V_{\mu\nu'\!,\!\nu\mu'} =\!\!\! \int\!\!
\psi^{e *}_{\mu}({\bf r}_1) \psi^{h}_{\nu'}({\bf r}_1)
 v({\bf r}_1 , {\bf r}_2 )
\psi^{h *}_{\nu}({\bf r}_2) \psi^{e}_{\mu'}({\bf r}_2) \
d{\bf r}_1 d{\bf r}_2
$$
where $v({\bf r}_1 , {\bf r}_2 )= |{\bf r}_1 -{\bf r}_2|^{-1}$ 
is the bare Coulomb potential while 
$w({\bf r}_1 , {\bf r}_2 )= \int \varepsilon^{-1} 
({\bf r}_1 ,{\bf r}) v({\bf r}, {\bf r}_2) d{\bf r} $ is the screened one. 
The excitonic eigenenergies $E_x$ and eigenstates ${\bm A}$
are provided by the solution of the above eigenvalue problem; hence,
this description \cite{hohe01prb} closely resembles  
the earlier one obtained within a Green's function scheme, 
which comprises the solution of the Bethe-Salpeter equation 
\cite{sham-rice66pr,hank-sham79prl}. 

The first step  
is therefore the calculation of
non-interacting electron and hole quasiparticle energies $\bar\epsilon$
and wavefunctions $\psi$ 
that we obtain through  a density-functional theory hamiltonian in the 
local density
approximation \cite{kohn-sham65pr,pwscf} 
(with norm-conserving pseudopotentials and a 
plane-wave basis with 50 Ry energy cutoff), and by subsequently adding a rigid 
shift to the conduction band energies to account for the self-energy correction 
(taken from Ref.~\onlinecite{rohl-loui99prl}).
Moreover, in order to make the full calculation computationally feasible 
for complex systems as polymer crystals, we 
approximate the full effective screening function
by a constant diagonal dielectric tensor 
${\underline{\underline{\varepsilon}}}$, whose components 
are calculated {\it ab initio} within linear response \cite{pwscf}. 
For the case of the isolated polymer 
chain, the assumption of space homogeneity is indeed very drastic, so we 
introduce an effective screening tensor, which is defined within the volume 
$\Omega_{\text{eff}}$  where the interaction actually takes place, while 
${\underline{\underline{\varepsilon}}} = 1$ outside $\Omega_{\text{eff}}$.  
The components of ${\underline{\underline{\varepsilon}}}$
are again calculated {\it ab initio} within linear response 
and $\Omega_{\text{eff}}$ is determined
selfconsistently with ${\underline{\underline{\varepsilon}}}$ to encompass the
region of significant electron-hole correlation function.

\begin{figure}
\centerline{\psfig{file=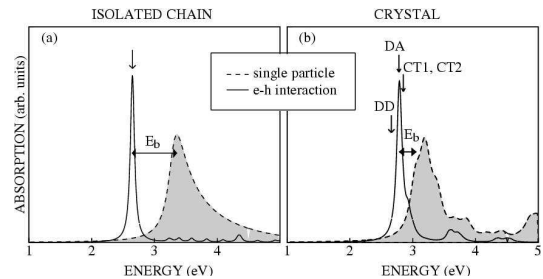,width=7cm}}
\caption{ $z$-polarized optical spectra for single chain PPV (left) and
the  PPV crystal in herring-bone structure (right).  The solid lines are
the {\em z}-polarized
spectra calculated including electron-hole
Coulomb interaction; the single particle spectra are shown for comparison
as shaded backgrounds. An artificial broadening of 0.04 eV (0.08 eV)
was included for plotting the Coulomb-correlated (single particle) spectra.
The exciton binding energy is the energy difference between the exciton
absorption peak in the correlated spectrum and the onset of the
continuum in the single-particle spectrum. The arrows mark the lowest
singlet exciton in the isolated chain spectrum, and
relevant singlet excited states in the
crystal spectrum: first optically active
direct exciton  (DA), optically inactive (dark) direct exciton (DD);
charge-transfer excitons (CT1 and CT2).} \label{spectra} \end{figure}
 
\begin{figure} 
 \centerline{\psfig{file=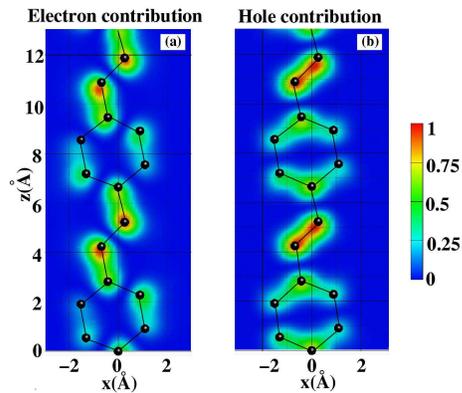,width=6cm}}
\caption{ Electron and hole contributions, $\rho^e$ (a) and $\rho^h$ (b), 
to the lowest singlet exciton  of the isolated PPV chain, plotted in a plane
parallel to the phenyl rings. The colour code is shown on the side bar, 
with $\rho^e$  ($\rho^h$) normalized to its maximum value.}
\label{eh-contrib} \end{figure}

In order to better understand the role of interchain coupling
for the crystal case, we first consider the isolated chain 
as reference system. 
We show in Fig.~\ref{spectra}(a) the results for the absorption
spectrum of the isolated chain \cite{rohl-loui99prl,ruin+01sm}, which
is highly polarized along the chain 
($z$-direction); for $x$ and $y$ polarizations
absorption close to band gap is negligible. We find an optically active
$ ^1 B_{u}$
singlet exciton with energy approximately 0.6$\div$0.7 eV below the
single-particle band-gap energy. 
This lowest singlet state extends over a few unit cells along the
chain \cite{rohl-loui99prl,ruin+01sm},
the large binding energy arising essentially from the strong lateral
confinement. To gain insight in the nature of this state, in 
Fig.~\ref{eh-contrib}
 we plot its electron and hole contributions: 
$\rho^{h} ({\bf r}_h ) = 
\sum_{\mu} | \sum_{\nu} A_{\mu\nu} \psi^{h}_{\nu} ({\bf r}_h )|^2$
and $\rho^{e} ({\bf r}_e) = \sum_{\nu}| \sum_{\mu} 
A_{\mu\nu} \psi^{e}_{\mu} ({\bf r}_e)|^2$.
Both have their maxima on the vinylene chains but in different bonding
configurations,  
and thus vinylene atoms are expected to undergo the largest
photoinduced lattice relaxation; this suggests the assignment of the
phonon replica observed in PPV photoluminescence to the vinylene C-C stretch
mode (around 0.2 eV \cite{tian+91jcp,saka+92jpc,mart+99prb,capa-cald02tobe}) and explains the
difference in lineshape \cite{eckh+89jcp} with poly-{\em para}-phenylene (PPP),
where the vinylene chains are not present.

We point out that the exciton signature is accompanied by suppression of the 
typical 1D van Hove  singularity ($\sim 1/\sqrt{E}$) \cite{ruin+01sm}. 
This is a phenomenon resulting from dimensionality, also found in inorganic 
semiconductor quantum wires \cite{ross-moli96all,ogaw-taka91prb}.
It is relevant for the experimental determination
of the exciton binding energy: the energy separation between the
exciton and the onset of the 1D absorption continuum cannot be extracted
directly from a one-photon optical spectrum, since the oscillator strength 
associated to the latter is vanishingly small \cite{ross-moli96all}; 
a separate measurement is required to extract the quasiparticle
gap \cite{leng+94prl,alva98-01all,camp+96prl,rina+01prb}.

\begin{figure}  \centerline{\psfig{file=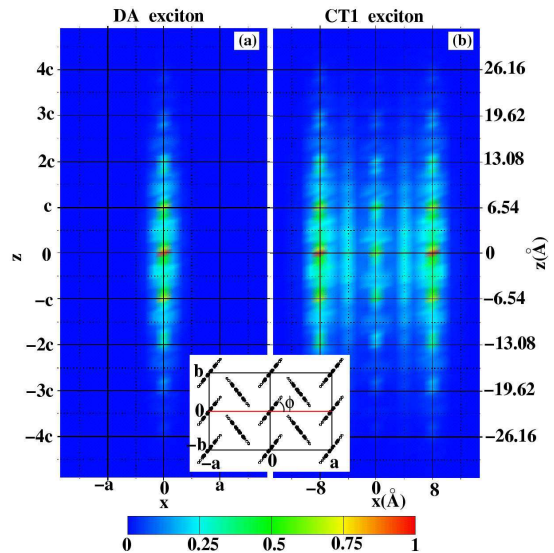,width=7.2cm}}
\caption{Probability distribution $|\Psi^{exc} ({\bf r}_e , {\bf r}_h)|^2$
of the lowest singlet excitons in the PPV crystal. 
The state (a) is the optically active direct 
exciton DA appearing in the spectra of Fig.~\ref{spectra}, while the state (b)
corresponds to the dark state CT1, at slightly higher energy. Here $\Psi^{exc}$
is plotted as a function of the electron-hole distance in the $xz$ plane
(red line in the inset) averaged over the center-of-mass
coordinate ${\bf R}=({\bf r}_e + {\bf r}_h )/2$. 
The colour code is shown below the figure,
with the probability normalized to its maximum value in each panel.
The inset shows a representation of the chains projection 
on the (a,b) plane of the orthorombic cell.}
\label{exci-wf} \end{figure}

We now turn to the PPV crystal, that we calculate in a realistic
herring-bone structure \cite{chen+92pol,gome-conw93prb},
by keeping an orthorombic cell with $a=8.07$ \AA, $b=5.08$ \AA, $c=6.54$ \AA{} 
and the setting angle $\Phi = 52^\circ$ (see inset in Fig.~\ref{exci-wf}).     
The influence of crystalline structure can be felt already on the
single-particle bands: due to the specific crystal symmetry, with
two chains per unit cell, all bands are doubled \cite{gome-conw93prb}; 
the splitting between doublets gives rise to the splitting between excitonic 
states (the so-called Davydov splitting
\cite{davy62}), and is strongly dependent on interchain interaction.
This feature alone points to a clear difference between PPV and
other substituted PPV derivatives, such as MEH-PPV, that cannot pack
in the herring-bone structure, have only one chain per cell,
 and thus would not show Davydov splittings. 
Note that  this very important feature
cannot be seen either from structureless crystal models \cite{hors+99prl}
or simple molecular aggregate models without 
translational symmetry \cite{corn+01sm}.
Other general features are that dispersion along the $\Gamma$-Z direction
(along the chain) is of the order of $\sim$ 2.2 eV, as for the isolated
chain, and dispersion perpendicular to the chain is vanishingly small. 

We show in Fig.~\ref{spectra}(b) the correlated spectrum for the
crystal. It is again characterized
by a strong singlet exciton peak (marked DA in the figure) and
quenching of the 
oscillator strength at the onset of the single-particle continuum.
While the single-particle band gap is reduced 
in the crystal \cite{gome-conw93prb},
the optical gap in the correlated spectrum is
close to the single-chain result. Hence, the binding energy of this exciton
is reduced to about 0.2 eV, consistent with the increased
screening in the crystal \cite{hors+99prl}. The first absorption peak is still
strongly polarized along {\em z}, confirming the strong optical
anisotropy of the material.

Now we turn our attention to the other remarkable features of the
optical behavior: (a) the single-particle
band-splitting results in an exciton Davydov splitting of $\sim$ 120 
meV, with the corresponding introduction of an optically
inactive (dark) state {\em below} the first active exciton (DD arrow
in the figure); (b) other
excitonic states are also introduced above the DA peak, with no
oscillator strength along {\em z}, but just barely active on the perpendicular
polarization (CT1 and CT2 in the figure).

The corresponding exciton wavefunctions  for DA and CT1 are shown in
Fig.~\ref{exci-wf},  where we plot the probability
density distributions $|\Psi^{exc} ({\bf r}_e , {\bf r}_h)|^2$.
Note that the exciton wavefunction $\Psi^{exc} ({\bf r}_e , {\bf r}_h)$,
expressing the probability amplitude of
finding the electron in ${\bf r}_e$ and the hole in ${\bf r}_h$,
is computed within our
formalism directly in terms of the interband polarization ${\bm A}$ as 
 $\Psi^{exc} ({\bf r}_e , {\bf r}_h ) = \sum_{\mu\nu}
A_{\mu\nu} \psi^{e}_{\mu} ({\bf r}_e ) \psi^{h*} _{\nu} ({\bf r}_h )$. 
The DA state is remarkably confined on a single chain, only slightly
more extended in $z$-direction than in the isolated chain. 
(The DD state has the same distribution as the DA, by symmetry: they are
both direct as electrons and holes are on the same chain.)  
The CT1 state, on the other hand, shows very small on-chain contribution,
with high probability of finding electron and hole on different
chains, and can be characterized as an excimer; same comments
apply to the CT2 state.

In summary, we have found that the lowest correlated states of PPV
crystal do involve more than one chain and are very dependent on
crystal structure \cite{buss+02tobe}; however, they retain a
quasi-one-dimensional character, and the 
consequent suppression of 1D singularity 
precludes the use of one-photon experiments
to extract exciton binding energies. 
Furthermore, we have found that, due to interchain coupling and symmetry 
effects, the lowest electronic excitation is a dark exciton, that is optically 
inactive but provides a non-radiative decay path. This can strongly quench the
photoluminescence with respect to non-interacting chains. 
In view of the impact of crystallization on the optical properties,
we studied also the influence of the packing density:
without changing the symmetry, we increased the interchain distance 
in the $\pi$-stack direction $b$ (see inset in Fig.~\ref{exci-wf});
we find that an increase of $\simeq$ 20 \% leads to an increase 
in the binding energy (to $\simeq 0.4$ eV) and a drastic decrease 
in the Davydov splitting by an order of magnitude,
thus restoring radiative efficiency.  
These results suggest that the spread in 
photoluminescence (PL) and photoconductivity (PC) data observed by
different groups may arise from differences in the aggregation state of the
polymer. Moreover, since real films have regions with different chain packings,
the dominant contributions to PL and PC can originate from different
regions of the same sample (the least and the most closely packed,
respectively).
Another solid state effect is the
appearance of bound charge-transfer excitations close above the first
exciton peak: this not only can enhance photoconductivity, but also can
be the origin of the infra-red active modes seen on two-photon
experiments \cite{mose+01sm_a,mose+00cpl}. 

Our results show that solid state effects are indeed very important
for the final optical structure of a polymer film. This conclusion
could only be reached because of  the possibility of calculating
first-principles optical spectra of realistic three-dimensional
polymer structures. 

We are grateful to Fausto Rossi for many relevant suggestions
in the early stages of this work.
We also thank Ulrich Hohenester, Lucia Reining, Eric K. Chang, 
 and Guido Goldoni for useful discussion.
This work was supported in part by MURST (Cofin-99) - Italy, by FAPESP -
Brazil, and by a CNR-CNPq bilateral project.


\end{document}